\documentclass[10pt, conference]{IEEEtran}
\IEEEoverridecommandlockouts
\usepackage[algoruled,linesnumbered,noline,noend]{algorithm2e}
\usepackage{mdframed}
\usepackage{xspace}
\usepackage{enumerate}
\usepackage[inline, shortlabels]{enumitem}
\usepackage{multirow}
\usepackage{tikz}
\usepackage{comment}
\usepackage{url}
\usepackage[T1]{fontenc}
\usepackage{amsmath,amssymb,amsfonts}
\usepackage{xcolor}
\usepackage{array}
\usepackage{booktabs}
\usepackage{cite}
\usetikzlibrary{shapes.geometric, shapes.misc, arrows,positioning, fpu}

\input{tex/contract-costs}

\SetKwProg{Fn}{Function}{ :}{}
\SetKwComment{Comment}{ $\triangleright$ }{}

\SetKwProg{Action}{On Action}{:}{}
\SetKwProg{Init}{Initialization}{:}{}

\newcommand{\Officials}{Officials\xspace}
\newcommand{\official}{official\xspace}
\newcommand{\officials}{officials\xspace}
\newcommand{\sdiploma}{credential\xspace}
\newcommand{\sdiplomas}{credentials\xspace}

\newcommand{\diploma}{academic credential\xspace}
\newcommand{\diplomas}{academic credentials\xspace}

\newcommand{\Diplomas}{Academic credentials\xspace}

\newcommand{\Certs}{Certificates\xspace}
\newcommand{\cert}{certificate\xspace}
\newcommand{\certs}{certificates\xspace}
\newcommand{\dcert}{public key certificate\xspace}
\newcommand{\dcerts}{public key certificates\xspace}
\newcommand{\contract}{smart contract\xspace}
\newcommand{\contracts}{smart contracts\xspace}
\newcommand{\blockchain}{blockchain\xspace}

\newcommand{\Blockchain}{Blockchain\xspace}

\def\BibTeX{{\rm B\kern-.05em{\sc i\kern-.025em b}\kern-.08em
  T\kern-.1667em\lower.7ex\hbox{E}\kern-.125emX}}

\begin{document}

\title{A Tree-based Construction for Verifiable Diplomas with Issuer Transparency}

\author{\IEEEauthorblockN{Rodrigo Q. Saramago}
    \IEEEauthorblockA{\textit{University of Stavanger}\\
        Stavanger, Norway \\
        rodrigo.saramago@uis.no}
    \and
    \IEEEauthorblockN{Leander Jehl}
    \IEEEauthorblockA{\textit{University of Stavanger}\\
        Stavanger, Norway \\
        leander.jehl@uis.no}
    \and
    \IEEEauthorblockN{Hein Meling}
    \IEEEauthorblockA{\textit{University of Stavanger}\\
        Stavanger, Norway \\
        hein.meling@uis.no}
    \and
    \IEEEauthorblockN{Vero Estrada-Galiñanes}
    \IEEEauthorblockA{\textit{University of Stavanger}\\
        Stavanger, Norway \\
        veg@ieee.org}
}

\maketitle


\begin{abstract}
    Still to this day, academic credentials are primarily paper-based, and the process to verify the authenticity of such documents is costly, time-consuming, and prone to human error and fraud. Digitally signed documents facilitate a cost-effective verification process. However, vulnerability to fraud remains due to reliance on centralized authorities that lack full transparency.

    In this paper, we present the mechanisms we designed to create secure and machine-verifiable academic credentials. Our protocol models a diploma as an evolving set of immutable credentials. The credentials are built as a tree-based data structure with linked time-stamping, where portions of credentials are distributed over a set of \contracts. Our design prevents fraud of diplomas and eases the detection of degree mills, while increasing the transparency and trust in the issuer's procedures.

    Our evaluation shows that our solution offers a certification system with strong cryptographic security and imposes a high level of transparency of the certification process.
    We achieve these benefits with acceptable costs compared to existing solutions that lack such transparency.
\end{abstract}



\section{Introduction}
\label{sec:intro}

\noindent
A \textit{\cert} is an official document attesting a certain fact~\cite{dictionary}.
In the education domain, \diplomas are seen as \certs that prove attainment of certain levels of academic merit of the bearer, which can confer advantages to individuals that hold such a \cert~\cite{CompetitionHigherEdu}.
Hence, such \certs are obvious targets for scammers seeking to exploit systems that depend on such documents.
Thus, there is an implied distrust between the issuer, \cert holder, and validator.

In the context of today's paper-based \diplomas, the processes involved in validating the authenticity of such \certs are notoriously slow and costly~\cite{FightCredFraud}, requiring significant human resources and expertise~\cite{BlockchainEdu}.
Once issued, a paper-based \sdiploma cannot be revoked, without involving the issuer in the validation process~\cite{BlockchainEdu}, which is rarely done~\cite{FightCredFraud}.
Furthermore, it may be difficult to distinguish \sdiplomas issued by legitimate institutions from those of degree mills, involved in the production of counterfeit \sdiplomas, made available for purchase~\cite{GlobalHigherEducationUnesco,DegreeMills}.
These challenges are further aggravated by the potential for bribery in the admission processes, as uncovered in Operation Varsity Blues~\cite{VarsityBlues}.

\Diplomas based on digital signatures have many advantages over paper-based ones; they resist tampering, verification can be automated, they are cheap to produce, can be revoked, and are expensive to forge~\cite{BlockchainEdu,cryptobook}.
However, \sdiplomas based on digital signatures would typically depend on \textit{certificate authorities}~(CAs), to issue and maintain a registry of valid \textit{\dcerts}~\cite{CA,cryptobook}.
Such a \dcert is used to certify the ownership of an entity's public key, whose corresponding private key can then be used to sign \sdiplomas.
To that end, CAs must be trusted by all parties involved in the certification process.
Hence, relying on CAs to control the certification and verification process may have serious security implications, since such central registries may be vulnerable to compromise, bribery, failure and censorship~\cite{BlockchainEdu,CertifiedLies,TrustDarknet,CertificateTransparency}.
Since \diplomas based on digital signatures do not hold intrinsic value without the registry, a failure in the registry would make the \sdiplomas worthless.
While CAs have been used to secure the web~\cite{CA}, several initiatives~\cite{CertificateTransparency,Convergence} have tried to address some of the concerns with CAs.
However, CAs have yet to see wide adoption for other applications.
One reason for this, is that there is no universally adopted standard for digital signatures, which often requires the use of specific software ecosystems for proper verification~\cite{BlockchainEdu}.

\Blockchain technology has emerged as a new database infrastructure that can provide public verifiability, integrity, and redundancy in a decentralized environment~\cite{DoYouNeedBlockchain}.
Hence, we can use a \blockchain as a reliable and immutable public notary for \certs, allowing anyone to perform certification and verification.
Further, records on the \blockchain cannot be modified without detection, and can only be lost if all copies on all peers in the \blockchain network are deleted.
However, existing ledger-based platforms for \diplomas have limited ability to detect or prevent corruption in the certification system~\cite{BlockchainUseInEdu}.
Some platforms~\cite{Stampery,EduCTX} are only accessible through proprietary interfaces or depend on a small number of trusted parties whose processes are neither transparent nor incorruptible.

Open standards like BlockCerts~\cite{BlockCerts} use the blockchain for timestamping certified \diplomas. 
Nevertheless, BlockCerts does not provide mechanisms for transparency of the issuing process, relying only on a single authorized signature to issue an authentic credential. 
The signer is susceptible to corruption; and if the signing keys get leaked, the security and credibility of the system will be compromised.

SPROOF~\cite{SPROOF} is another platform for issuing, managing, and verifying credentials on a permissionless \blockchain.
In SPROOF, credentials are created using a key derivation function similar to hierarchical deterministic wallets~\cite{hdwallets}.
Subjects allow the issuer, by sharing a master seed key, to derive new sub-keys and the corresponding hash of granted credentials.
Such a mechanism forms a pseudonym tree, i.e., tree of public keys, with derived keys and credentials in the leaves, allowing selective disclosure of credentials under the sub-trees of the shared key.
However, as in BlockCerts, issuers are represented by a single key-pair.
Thus the system is also vulnerable to corruption and a single point of failure.
Moreover, verifiers need to iterate over all transactions in the \blockchain sent by the SPROOF address to verify the credentials.

This paper presents a novel document verification protocol designed to strengthen the trustworthiness and transparency of certification systems.
We accomplished this by using ledger-based timestamping of individual certification steps that are composed into a final certificate.
To prevent that only a few \officials can compromise the final certificate, we distribute the individual certification steps among ``all'' the \officials involved in issuing the certificate.
Hence, bribing only a few \officials is insufficient to compromise the final certificate.

Our protocol is based on a tree structure of \contracts that \officials interact with to certify their assigned part of the final certificate.
\Officials are granted permission to certify their assigned part following a hierarchical structure.

We apply our protocol to \diplomas, where each certification step is performed by one or more faculty members acting as \officials.
This approach fits well with existing procedures, where faculty members submit grades for inclusion in the \diploma.
Hence, our protocol reduces the power of each \official to a single certification step.

We implement our construction using \contracts and evaluate two alternative deployments.
One is on a public permissionless blockchain based on Ethereum.
The other is a consortium-style hybrid deployment using permissioned and permissionless blockchains.
In this deployment, the \contracts mainly operate in the permissioned realm, with periodic anchors recorded on the public permissionless blockchain.

The first approach has the benefit of strong security, transparency, and infrastructure reliability, which could enable the creation of a global system for digital credentials in practice.
However, it is currently throughput-limited and suffers a significant disadvantage in terms of cost with today's gas prices and Ether exchange rates, although it may change when Ethereum completes its migration to proof-of-stake.
While our second approach requires that the consortium maintains the infrastructure. While this requirement may reduce reliability, it offers an appealing tradeoff between cost and security.


\section{Background}
\label{sec:bg}

\subsection{The Certification System}
\label{sec:cert-system}
\noindent
A \cert is a document confirming a statement about a subject, e.g., Lamport has a degree in Mathematics. \Certs can be issued by entities with authority to do so.

The objective of a certification system is to prevent \mbox{invalid} \certs, while ensuring correct \certs are widely accepted by third-party verifiers.
A \cert is considered valid if these criteria are satisfied:
\begin{enumerate}[(a)]
    \item \label{crit:authenticity} It was issued by the authority and for the subject claimed.
    \item \label{crit:integrity} It always represents the same claim as when issued. 
    \item \label{crit:issuerstandard} The issuance is a standardized procedure. 
    \item \label{crit:accreditation} The issuer's procedure meets external standards. 
\end{enumerate}

Criterion~\ref{crit:issuerstandard} is typically assumed correct or only performed by the issuer's \officials, and trust in the issuer is assumed~\cite{BlockCerts}.
Criterion~\ref{crit:authenticity} and \ref{crit:integrity} commonly make use of digital signatures and provide mechanisms to identify the entities.
These mechanisms frequently depend on the involvement of third-parties and are verified using identity credentials, which themselves are \certs attesting to a person's identity~\cite{BlockchainEdu}.

Similarly, \ref{crit:accreditation} may be verified by some regulatory agency when it has the authorization to examine the issued credentials.
For the \diplomas case, \ref{crit:accreditation} is often checked by reviewing the courses taken as part of a degree.

\subsection{Digital Certificates}
\label{sec:digital-certs}

\noindent
Digital signatures provide more powerful capabilities than their paper-based counterparts.
They are easy to create and verify, and they are secure as long as the private key is kept secret.
However, using digital signatures to authenticate credentials gives rise to two problems:
\begin{enumerate*}[(1)]
    \item \label{problem:PKI} a verifier needs to know that a certain key indeed belongs to the issuing authority;
    \item \label{problem:keytheft} a compromised private key may allow arbitrary creation of authentic credentials.
\end{enumerate*}

Problem~\ref{problem:PKI} is similar to the paper-based version, where signature verification requires trusted samples, and is usually solved using CAs.
The typical approach to address~\ref{problem:keytheft}, is to issue credentials with a limited expiration time.
This allows a change of keys when reissuing credentials.
However, frequent reissuing of \diplomas is not feasible, since they are expected to have life-long validity and may be used rarely.

In the era of cloud computing, an alternative is to use an online data registry, that can process verification requests.
This approach allows revocation of \certs, which in turn helps to mitigate the problem of compromised keys.
The problem with this approach however, is that the data registry may form a single point of failure.
For life-long \certs, data loss at the registry would be detrimental.
Additionally, tampering with the data stored at the registry may be difficult to detect.

For these reasons, there is no broadly adopted certification system for \diplomas based on digital signatures.

\subsection{Blockchain Data Registry}
\label{sec:blockchain-registry}

\noindent
A \blockchain system further expands on the capabilities of digital signatures, and is typically implemented as a geo-replicated service that tolerates arbitrary failure and compromise of a fraction of the replicas.
Correct replicas maintain the complete state of the \blockchain and process the same updates.
What characterizes a \blockchain is that updates are written to a crypto\-graphic\-ally immutable data structure.
This immutability property comes from the combination of replicating its blocks on all participants, and cryptographically linking the blocks.
This property makes it infeasible to modify previously written data, without the collusion of a majority of the replicas.
A \blockchain system may also execute arbitrary code, i.e.,~\contracts, as part of a state transition~\cite{Ethereum}.
Finally, a \blockchain can also be used to implement a trusted timestamping service~\cite{OriginStamp,TTS,PoE}.

Taken together, the above capabilities makes a \blockchain a suitable structure for implementing data registries for digital \certs.
For instance, BlockCerts~\cite{BlockCerts} uses the \blockchain as a tamper-resistant data registry for timestamped \diplomas.
This is achieved by publishing the hash of \sdiplomas on either Bitcoin or Ethereum, through a transaction signed by the issuer.
BlockCerts can also revoke \sdiplomas.
However, BlockCerts is susceptible to compromise and misuse via the issuer's private key used to sign \sdiplomas, potentially violating~\ref{crit:issuerstandard}.
Furthermore, the revocation process in BlockCerts relies on central servers hosted by the issuers.
Other systems suffer from the same problem~\cite{BlockNotary,EduCTX,Registree,Stampery}.


\section{System Model}
\label{sec:model}

\noindent
We now present our system model, following the W3C data model for \textit{verifiable credentials}~\cite{W3C-VCDataModel}, with the addition of the \textit{registrar} role that is a critical role performed by one or more \officials that are members of the issuer organization, as explained in Section~\ref{sec:responsibility}.
In this model, a credential holds data in the form of claims and metadata necessary to verify the credential's authenticity.
We consider a certification ecosystem composed of entities, performing one or more roles in the system, and interacting with each other to establish a trustful relation around the exchanged credentials.
The entities can play the following roles in the system:

\begin{table}[h]
    \centering
    \def\arraystretch{1.1}
    \begin{tabular}{@{\hspace{0.2em}}p{.15\linewidth}@{\hspace{0.2em}}|@{\hspace{0.2em}}p{.8\linewidth}}
        \toprule
        \textbf{Subject}                  & Entity about which claims are made, e.g., a student.                                                                                                                       \\
        \textbf{Holder}                   & Entity that possesses and presents \textit{verifiable credentials} or any data derived from it.                                     \\
        \textbf{Issuer}                   & Entity that executes the certification process, by asserting claims about subjects and creating \textit{verifiable credentials} from these claims, e.g., universities.     \\
        \textbf{Registrar}                & Entity that executes actions of the certification process on behalf of an issuer.                                                                                         \\
        \textbf{Verifier}                 & Entity that evaluates whether a verifiable credential is authentic and valid in a given context, e.g., employers.                                                          \\
        \textbf{Verifiable Data Registry} & A certification system that mediates the creation and verification of keys, identifiers, and \textit{verifiable credentials}, e.g a trusted database, distributed ledger. \\ \bottomrule
    \end{tabular}
\end{table}

\noindent
For simplicity, we consider the subject to be the same as the holder, and we use the terms interchangeably.

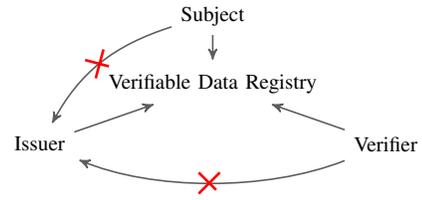
\begin{figure}[h]
    \centering
    \begin{tikzpicture}[->, >=stealth', shorten >=2pt, auto, node distance=1.5cm]
        \tikzset{cross/.style={cross out, draw=red, line width=1,
                    minimum size=2*(#1-\pgflinewidth),
                    inner sep=0pt, outer sep=0pt}}
        \tikzstyle{state} = [draw=none]
        \tikzstyle{arrow} = [semithick, draw=black!64]

        \node (subject) [state] {\footnotesize{Subject}};
        \node (vdr) [state, below of=subject, yshift=0.6cm] {\footnotesize{Verifiable Data Registry}};
        \node (issuer) [state, left of=vdr, xshift=-0.8cm, yshift=-0.8cm] {\footnotesize{Issuer}};
        \node (verifier) [state, right of=vdr, xshift=0.8cm, yshift=-0.8cm] {\footnotesize{Verifier}};

        \draw [arrow] (subject) to (vdr);
        \draw [arrow] (issuer) to (vdr);
        \draw [arrow] (verifier) to (vdr);
        \draw [arrow] (verifier) to[bend left=22,looseness=0.8] node[cross=5pt,anchor=text]{} (issuer);
        \draw [arrow] (subject) to[bend right=22] node[cross=5pt,rotate=30,anchor=text]{} (issuer);
    \end{tikzpicture}
    \caption{Trust relation between the roles in a certification ecosystem~\cite{W3C-VCDataModel}. Our model removes the trust requirements between the issuer, the subject, and the verifier by adding transparency in the certification process.}
    \label{fig:trust-relation}
\end{figure}

\figurename~\ref{fig:trust-relation} shows the trust relationship between the different roles~\cite{W3C-VCDataModel}. For each role, the arrows show, which other roles it needs to trust.
The subject, the verifier, and the issuer need to trust the verifiable data registry to provide correct information, and at the same time, the subject and verifier need to trust the issuer to create credentials that assert the truth about the subject~\cite{W3C-VCDataModel}.

In this work we use \blockchain technology to implement both a trustworthy verifiable data registry and trustworthy issuers without relying on external trust assumptions.
We assume that each entity is represented by one or more cryptographic key pairs used to sign and verify messages, and that cryptographic primitives cannot be trivially circumvented.

\subsection{Trustworthy Data Registry}

\noindent
Issuers, subjects, and verifiers trust the data registry to keep tamper-proof data, ensure its availability, and enforce access control on write operations.
We justify this trust through the use of \blockchain technology.
We note that both permission\-less and permissioned \blockchain systems can be used, provided that the consortium of the latter is sufficient to justify trust from third parties.

We assume that the \blockchain allows write access to issuers and subjects, and read access to verifiers.
Additionally, we assume a \blockchain that enables the deployment of \contracts, and that the blockchain's setup and mechanics are sufficient to justify trust among participants, including third-party verifiers.

\subsection{Trustworthy Issuers}
\label{sec:type-frauds}

\noindent
According to W3C's trust model~\cite{W3C-VCDataModel} holders and verifiers need to trust the issuer to issue valid \sdiplomas about subjects and to revoke them when necessary.
To grant this amount of trust to an issuer, the following attacks must be prevented:

\begin{enumerate}
    \item Tampering: Unauthorized content manipulation of valid issued \sdiplomas, including by authorized personnel.\label{threat:malicious-edit}
    \item Compromised infrastructure: Unauthorized creation of \sdiplomas due to compromised private keys etc.\label{threat:fake-creds}
    \item Creation of fake \sdiplomas by
          \begin{enumerate*}
              \item Unaccredited issuers.\label{threat:corrupt-issuers}
              \item Impersonation of legitimate issuers.\label{threat:mimic-issuers}
          \end{enumerate*}
\end{enumerate}

\noindent
We refer to the issuer as \emph{fraudulent} in all the above cases.
Tampering can be prevented through cryptographic primitives.
To address the threat of compromised infrastructure, our certification system explicitly models different registrars acting on behalf of the issuer and ensures that even in the presence of some corrupted registrars,
\certs are issued according to standardized procedures (Criterion~\ref{crit:issuerstandard}), that can be validated (Criterion~\ref{crit:accreditation}).
While cryptographic primitives may prevent impersonation, our system must rely on an accreditation agency to certify and compare different issuer's procedures.
For instance, in the case of \diplomas, trusted issuers may still certify graduates with various skill levels.
Nevertheless, even without accreditation, the possible actions of fraudulent issuers can be significantly limited or flagged as suspicious, as we discuss in Section~\ref{sec:disc}.


\section{Security Mechanisms}
\label{sec:mechanisms}

\noindent
We now define the abstract mechanisms of the certification system that helps to detect and prevent fraudulent behaviors.

\subsection{Credential Composability}
\label{sec:composition}

\noindent
Our first mechanism is based on the realization that the process leading towards the issuance of a credential can be modeled as several sub-credentials, issued sequentially over a period of time, and representing an accumulation of real-world achievements.
Hence, we propose to record such sub-credentials and compose them into a final \emph{composed credential}.

The resulting dependency graph enables verifiers to perform automatic checks and detect malicious actions in the certification process and establish a causal relation between achievements that can be publicly verified.

This approach differs from existing certification systems that assume trust in the issuer's actions and omit information about how credentials are created.
Such omission can be critical for longer processes, e.g., for a bachelor's degree, resulting in creating only one final credential, which a fraudulent issuer's action could have created.

\subsection{Trusted Timestamp}
\label{sec:timestamp}

\noindent
Trusted timestamping~\cite{TTS} is a method for associating data with time.
To that end, we leverage the \blockchain to establish trusted time information for all data stored on the chain.
The use of trusted timestamping prevents even authenticated issuers from back-dating credentials, and is thus a powerful primitive that can limit fraudulent actions.
Trusted timestamping is especially useful in combination with credential composition, since timestamps allow the verification of the timeframe, temporal distribution, and succession of the sub-credentials.
By timestamping each sub-credential, verifiers can check whether the entire set of credentials of a subject, represents a consistent and ordered sequence of his achievements.
Further, the time information can also be used to identify bogus patterns in a certification process.

\subsection{Distributed Responsibility}
\label{sec:responsibility}

\noindent
The authenticity of digital credentials depends on the possession of certified keys to sign credentials.
If such keys are compromised or misused by authorized individuals, it could lead to severe exploitation.

One approach to prevent a single compromised key from compromising the whole system is to employ a multi-signature scheme, where multiple certified keys are needed to create a credential.
For additional security, the keys can be kept separate and used by different registrars.
However, this approach is limited by its high administrative cost.

Our credential composition mechanism offers a different approach, where the responsibility to issue sub-credentials can be distributed among different registrars over distinct steps of the process.
The use of trusted timestamping allows assigning a registrar's responsibility to a limited and specific time-period.
This reduces the administrative overhead of multiple signers, making it more likely that registrars can detect fraud.

Creating a complete credential in our model depends on the actions of multiple registrars in different steps of the certification process.
Each step validates the previous steps on which they depend, and are potentially performed and signed by different registrars.
By subdividing credential creation into multiple steps, we separate the responsibility of issuing credentials, enabling accountability of the certification procedures.


\section{Certification Protocol}
\label{sec:credential-construction}

\noindent
In this section, we present the protocol that implements the mechanisms described in Section~\ref{sec:mechanisms}.
The protocol builds on a \blockchain that supports \contracts.
The \blockchain acts as a verifiable data registry for certification processes.

For deploying the \contracts, we consider two alternative \blockchain architectures.
First, in a permissionless \blockchain architecture, all storage and \contract execution is conducted on a public ledger.
Second, we also consider a hybrid permissioned and permissionless architecture, where the majority of storage and \contract execution is conducted on a private consortium ledger, with periodic anchors recorded on the public ledger.
In the following, we describe the certification processes running in the hybrid architecture.
We discuss the pros and cons of these two alternative designs in Section~\ref{sec:eval}.

\subsection{Trustful Certification Process}
\label{sec:cert-tree}

\noindent
Let an \emph{issuer} be any organization that produces verifiable credentials in some certification ecosystem.
In our protocol, we model the issuer as a set of \contracts, called \emph{notaries}.
These notaries define the certification process for issuing certificates from the organization.

The notaries run in a permissioned \blockchain controlled and maintained by the organization. 
A certification process may be composed of several \emph{steps}, each represented by a notary.
A notary has one or more registrars authorized to perform certification steps on behalf of the issuer.
Registrars are assigned to a notary on deployment based on their responsibility for some step in the process.
Further, a notary encodes rules required to complete a certification step, e.g., at least two registrars must sign a document before it is considered valid.

The assignment of registrars to notaries typically reflects the issuer's organizational structure.
Thus, in our protocol, notaries are nodes in a tree data structure that we call a \emph{certification tree}.
A certification tree establishes the hierarchy of responsibilities in the certification system and implements the mechanism described in Section~\ref{sec:responsibility}.

The certification tree is created top-down during the deployment of the initial set of notaries.
The tree grows over time by adding new notaries as children of existing tree nodes.

The notaries in the certification tree can perform three operations:
(1)~add a node to the tree, (2)~register subjects, and (3)~manage credentials.
Operation~(1) allows a notary to create a new node in the tree as long as the parent notary's rules are satisfied.
The parent notary can delegate authority to the added notary's registrars to perform certification steps.
Registrars can only add nodes beneath the tree node they are authorized to extend.
Operation~(2) allows a notary to register new subjects in the system to certify them.
This operation may require identity checks, which would naturally be part of an enrollment process.
Operation~(3) gives the ability of a notary to issue or revoke credentials.

A \emph{leaf notary} does not have descendants and represents steps of the certification process that do not depend on other steps.
Leaves cannot add new nodes to the tree.
A \emph{non-leaf notary} represents steps that depend on the certification steps of their descendants.
Further, a non-leaf notary can issue credentials based on evidence provided by a subset of its descendant notaries, also called \emph{witness} notaries.
The evidence is based on the credentials already issued by witness notaries for a designated subject.
Thus, a non-leaf notary acts as an aggregator of credentials issued by its descendants.

Since leaf notaries do not have witnesses, credentials issued by leaves must be authenticated by their respective registrars.
Credentials registered in the lower levels of the certification tree facilitate creating credentials in higher levels.
Notaries may require multiple steps to complete a certification process, resulting in a verifiable credential in each step.

\begin{figure}[t]
      \centering
      \includegraphics[width=\linewidth]{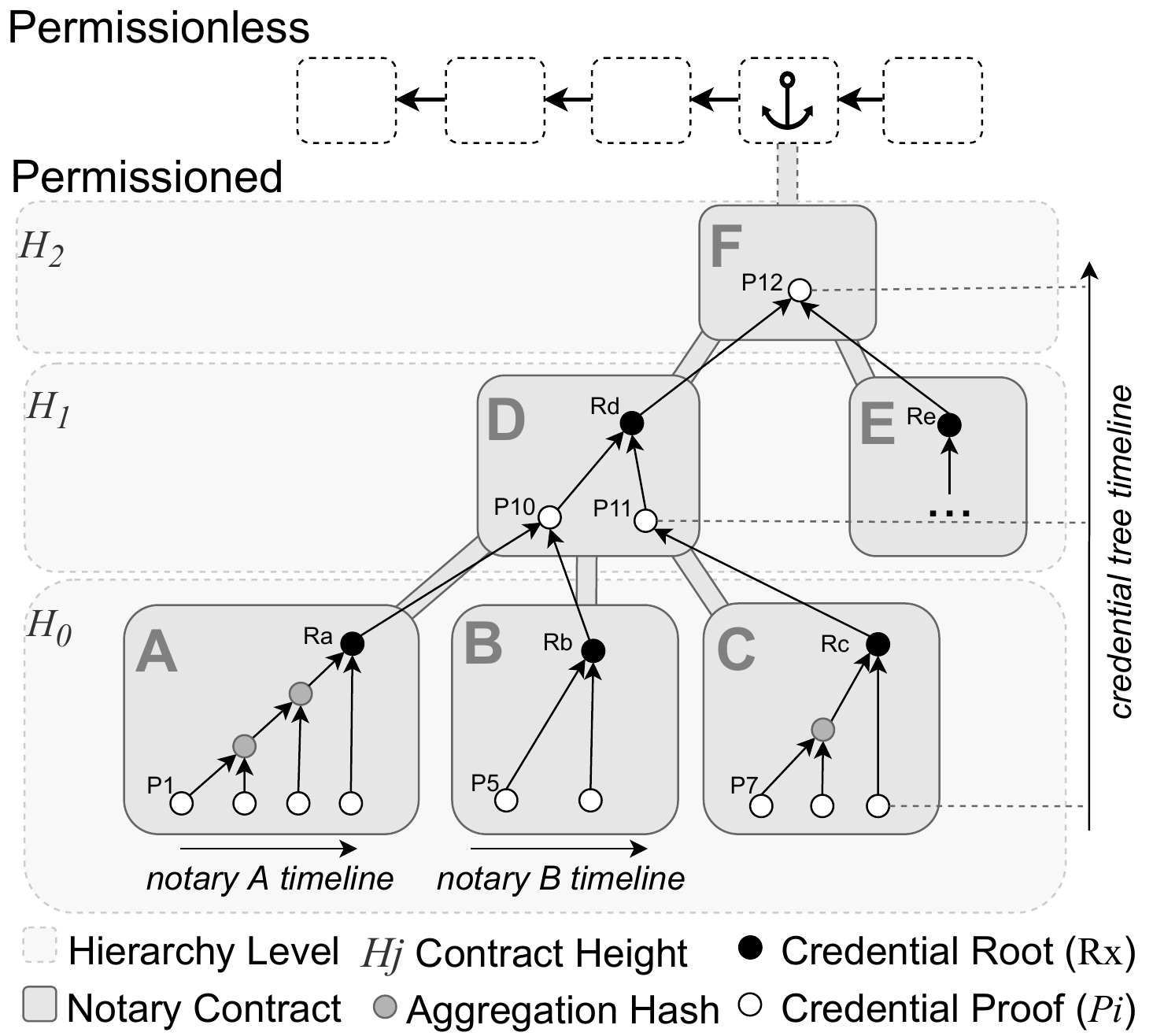}
      \caption{Certification tree and the credential tree over the \emph{credential proofs} for a subject. $R_{x}$ denotes the credential root aggregated from all credentials of a subject in contract $x$, where $x \in \{A,B,C,D,E,F\}$. $H_{j}$ denotes the height $j$ of the contract in the certification tree. $P_{i}$ represents a credential proof. Notary $F$ is anchored to another notary in a permissionless blockchain.}
      \label{fig:cert-tree}
      \vspace{-0.5cm}
\end{figure}

\figurename~\ref{fig:cert-tree} shows an example certification tree, in which the gray rounded squares are notary contracts labeled $A,B,C,D,E,F$.
In the example, notary $D$ has issued credential $P_{10}$, justified by credentials from notaries $A$ and $B$.

Notaries operate over \emph{credential proofs}, i.e., public evidence of the issuance or revocation of a verifiable credential.
The proof contains metadata about the credential and is stored in the \contract state.

The holder stores the credential itself.

When a credential is to be issued, a registrar generates a credential proof and registers this in the notary contract's state.
Following this, a quorum of the notary's registrars and the credential's subject start an agreement period in which they must accept or reject the registered credential.
Traces of this agreement process can be audited by anyone with read access to the \blockchain.

A quorum of signatures of registrars defined during the notary's deployment is required to authorize the notary's procedures.
Each credential proof has information about the quorum that authorized its creation.
Although notaries do not include any business logic, issuers may have different constraints for authorizations and can implement such logic on top of notaries.

\subsection{Constructing Credentials}
\noindent
Credentials are created by the certification tree using the mechanism described in Section~\ref{sec:composition}, and every credential proof has information about the witnesses that justified its creation unless a leaf created it.

We obtain another tree-like data structure called a \emph{credential tree} by applying a progressive composition of previously issued credentials across the certification tree.
The credential tree defines a causal set of events that enable issuing the credential at higher levels of the certification tree.
Each notary may contain multiple credential trees, one per subject.

Credential trees are created bottom-up by the registrars.
A certification process begins in a leaf of the certification tree, expanding to the parent nodes until it reaches a defined height or the root notary in the certification tree.
Thus, a credential tree is a rooted tree in which its vertex set is given by the credential proofs in a subset of notaries of the certification tree.
\figurename~\ref{fig:cert-tree} also shows an example of a credential tree, where the black and white dots shows links between the notaries.

We consider a credential to be \emph{authentic} if a credential proof exists in the \blockchain, such that the proof was signed by a \textit{quorum} of registrars and \textit{approved} by the subject.
Further, an authentic credential is composed of a partially ordered set of credential proofs distributed over the notary contract's state in the certification tree.

The \emph{credential root} is the root of a credential tree created by a notary when aggregating all credential proofs of a subject, according to their insertion order in the \contract.
The credential root defines a point in the certification process achieved by the subject at a specific height of the certification tree. An ancestor node uses it as evidence to support the creation of credentials in the parent.
When a notary's credential root is used as evidence to create a credential in a parent, we say that such a notary is a witness to the credential.

Disjoint credential subtrees are unified by issuing a credential proof on the parent contract using the subtree notaries as witnesses.
However, all credential subtrees must be for the same subject and have a common ancestor.

Finally, time information provided by the \blockchain is encoded into each credential proof registered by the notary, establishing the trusted timestamp mechanism described in Section~\ref{sec:timestamp}.
Hence, the credential proof cannot be changed without evidence of tampering, even by the issuer's authorities.

\subsection{Anchoring Certificates}
\label{sec:anchor}

\noindent
Let \emph{anchor} be a \contract deployed in a permissionless \blockchain.
The anchor contract is responsible for periodically publishing credential roots resulting from the certification processes performed by the notaries in the certification tree.
The frequency by which the anchor can be updated with a new root is configurable and depends on the issuer's requirements.
For the academic scenario described in Section~\ref{sec:impl}, a typical interval could be every semester.

The anchor contract serves as a trusted on-chain entity representing the issuer on a public permissionless \blockchain.
Such anchor allows to extend the security, i.e., timestamping and immutability, and availability guarantees from the permissionless \blockchain to credentials issued on a permissioned \blockchain but reduces the amount of data that needs to be stored on the public chain.
It also enables interoperability between different issuer platforms that implement our protocol.


\section{Security Analysis}
\label{sec:disc}

\noindent
This section discusses how our protocol described in Section~\ref{sec:credential-construction} addresses the threats described in Section~\ref{sec:type-frauds}, here denoted as \textit{Frauds}.
We demonstrate that our protocol can limit the damage in the certification system due to fraudulent issuers, improving the trust in the whole system while inhibiting the action of malicious entities.

\subsection{Certification Fraud}
\label{sec:cert-fraud}

\noindent
We consider \emph{certification fraud} to be an attempt to make illegitimate credentials by tampering with existing ones or creating credentials from scratch for unqualified subjects.
Such attacks may be conducted by authorized registrars, e.g., through bribery or by external agents that have gained access to the issuer's infrastructure or a registrar's private keys.
In the following, we explore various attacks and how our protocol addresses them.

\subsubsection{Tampering}
\label{sec:tampering-cred}
The blockchain's irreversibility guarantees and the use of document digests prevent tampering with issued documents, both by the subject, the issuer, or a third party. These measures ensure Criterion~\ref{crit:integrity} and prevent Fraud~\ref{threat:malicious-edit}.

\subsubsection{Compromised Infrastructure}
\label{sec:corrupt}

Our protocol limits the attack surface due to stolen keys (Fraud~\ref{threat:fake-creds}).
If keys with the power to create new notaries, issue/revoke credentials or assign registrars are compromised, the distribution of responsibilities under the certification tree may not be effective against fraud.
To prevent such attacks, we require a quorum of signatures for these critical operations, minimizing the impact of compromising individual keys.

While a corrupt registrar may create a fraudulent credential on a leaf notary, creating a credential tree requires multiple notaries and possession of matching signing keys.
To discourage such attacks, the enrollment process that assigns roles to registrars is recorded in the credential tree for accountability.

Further, we note that we can easily detect certification subtrees created with too few registrars during validation, identifying potential malicious behaviors. 
Even if an attacker succeeded in deploying a fraudulent subtree, this subtree would be disjoint from valid credentials due to the hierarchical construction.
Thus it is easy to detect and revoke all credentials created based on this subtree when detected.

\subsubsection{Preventing Credential Factories}
\label{sec:false-issuers}

To create valid credentials using our protocol, an issuer must create a valid credential tree distributed over some realistic time-period, e.g., three years for a credential tree of a bachelor's degree.
This time restriction severely limits the practicality of unaccredited issuers, such as degree mills (Fraud~\ref{threat:corrupt-issuers}), since their customers would have to wait for the same time-period.
Further, using a public data registry makes it easy to detect fake issuers and expose them publicly.

Furthermore, authentic issuer organizations would typically be approved by national accreditation agencies.
While the accreditation process may suffer from similar fraud risks, our protocol can also model the accreditation process.
Hence, in the context of a consortium network of educational institutions deployed on a permissioned \blockchain, only accredited issuers would be allowed to join.



\subsubsection{Illegitimate Credentials}
\label{sec:illegitimate-cred}

Storing only the hash of the final credential document, as done in BlockCerts, reveals nothing about the process required to obtain the credential, which is essential to prevent or detect a fraudulent certification process.
That is, a registrar would be authorized to issue credentials to anyone in the system.
Thus, the actions of a registrar should be restricted and monitored, preferably by an entity that cannot easily be bribed or manipulated.
In our protocol, we use \contracts that encode rules that can impose various restrictions and enable accountability of the registrar's actions.
Furthermore, we distribute the responsibility of issuing sub-credentials to many registrars, increasing the overall trust in the certification process.

\subsection{Impersonation Attacks}
\label{sec:id-fraud}

\noindent
Our protocol can defend against Fraud~\ref{threat:mimic-issuers}, i.e., impersonation attacks.
Below, we consider both issuer impersonation and partial impersonation via an issuer's agents.
Digital signatures on their own do not protect against impersonation without a public key registry certifying which key belongs to which entity~\cite{BlockchainEdu}.
Henceforth, we assume the existence of such a registry~\cite{CA,DPKI} for certifying issuers and other entities in the system.
In Ethereum, the Ethereum Name Service~(ENS)~\cite{ENS} may also certify such entities.

\subsubsection{Impersonating Issuers}

In our protocol, a verifier can check the issuer's identity against the \contract at the root of the certification tree.
Hence, if a well-known domain name for the issuer is recorded in the ENS name service and maps to the issuer's root contract, a verifier can perform such a check by inspection.
Thus, no other issuer can point to the same contract, and it cannot be impersonated.

Even in the absence of such registries, impersonating issuers would still be restricted via the accreditation process, as described in Section~\ref{sec:false-issuers}.

\subsubsection{Impersonating Agents}


Our protocol is based on the verifiable credential data model~\cite{W3C-VCDataModel}.
It allows using any supported identification method, e.g., a decentralized identifier~\cite{W3C-DID}, to prevent impersonation.
The following mechanism is used to add new agents to the system, i.e., both registrars and subjects.

The identifier of an agent is included in the credentials tree construction, thus protecting it from tampering.
With this construction, we can create a composable enrollment process linking to an agent’s identity.
Clearly, this enrollment process requires physical verification of the agent's identity.
The agent's enrollment credential can be extended with the agent's credential tree throughout the certification process, thereby linking the agent's identity to the credential tree.

\subsection{Revoking Credentials}

\noindent
We allow each notary to revoke credentials it has created by adding them to a revocation list stored in the notary contract. 
Thus a notary in a higher level of the certification tree can revoke an aggregated credential, but not the credentials issued by its descendants.
We allow both registrars and subjects to revoke credentials but recommend using quorums with multiple registrars on the revocation of aggregated credentials.

\subsection{Privacy Considerations}

\noindent
In this section, we discuss various aspects related to information stored in credentials and the privacy of this information.

In real-world scenarios, it may be desirable for an entity to supply identifying information and credentials to third parties to establish a trust relationship.
For instance, a student's data can be helpful for potential employers when the student is applying for a job.
Similar for loan applications or admission processes to a higher education program.

Note that a third party cannot manipulate a subject's credentials or perform actions in the subject's name in our protocol.
However, it is still possible to prove the subject's real-world identity with on-chain actions through identification services.
That is, the subject can prove control over the public key used to sign the relevant credentials simply by performing a transaction using the key.


Once issued to a subject, our model allows credentials to be under the control of the subject while still ensuring its authenticity and correctness.
Anyone can verify a credential without authorization from the issuer.
Verifiers can inspect the actions of the notaries under the certification trees and check the content of the credentials disclosed by the subject without revealing more information than necessary.

This work follows the W3C data model specification~\cite{W3C-VCDataModel}, with credentials stored as JSON documents, as described by the specification.
Such credentials can also be created using \textit{Precise-Proofs}~\cite{Centrifuge}, allowing subjects to disclose specific parts of a credential with authenticity guarantees.

Knowing a subject’s public key or address makes it possible to identify all credential proofs of the subject.
Such a link between credentials is required in our system to allow on-chain validation of the issuing process without depending on the issuer's approval or third-party attesters.
This can potentially limit the scope of the proposed protocol to use cases where such public links are acceptable, e.g., the education scenario.
We plan to address this limitation in future work.

Despite the possibility to find a subject's credential proofs, no meaningful information about the credentials is exposed, as long as the holder's chosen storage system is secure and encryption keys are not compromised.



\section{Case Study: Academic Scenario}
\label{sec:impl}

\noindent
In this section, we describe how we implement a document verification system for \diplomas using the protocol presented in Section~\ref{sec:credential-construction}.
\Diplomas are represented as a collection of individual exam \sdiplomas, each attesting to a student's achievement during some study period.

We implemented our protocol as an open-source library\footnote{https://github.com/relab/certree} using the Solidity programming language on top of the Ethereum \blockchain.
Our implementation emulates the certification processes of hypothetical universities and divides the responsibilities for creating a diploma between different entities at a university.
Students and employees of a university are represented by their Ethereum accounts, and notaries are represented by \contract accounts forming the certification tree described in Section~\ref{sec:credential-construction}.

Each university has its own certification tree and an anchor \contract, as described in Section~\ref{sec:anchor}.
The root of the certification tree, i.e., root contract, is registered in the anchor contract.
The anchor contract represents the university's identity for entities outside the system, and it is mapped to the university's well-known domain name, using the ENS~\cite{ENS}.

Further, we extended our protocol by implementing the course, department, and university operations as \contracts that use the notary functionality as a library. The source code of this extension is also available\footnote{https://github.com/relab/credcontracts}.
These contracts are managed by the university's employees and compose the nodes of the university's certification tree.

Consider again the certification tree shown in \figurename~\ref{fig:cert-tree}, but now from the university's perspective.
Notary $F$ represents the university's root contract and has a set of department contracts as children, i.e.,~notaries $D$ and $E$.
The departments are created by members of notary $F$ and are managed by department employees, typically administrative staff, i.e., registrars.
The leaves $A$, $B$, and $C$ are course contracts created by members of the department.
Course contracts are managed by a set of evaluators and instructors, specified at the time of deployment of the course.

\subsection{Issuing Academic Credentials}
\label{sec:issuing-cred}

\begin{figure}[t]
      \includegraphics[width=\linewidth]{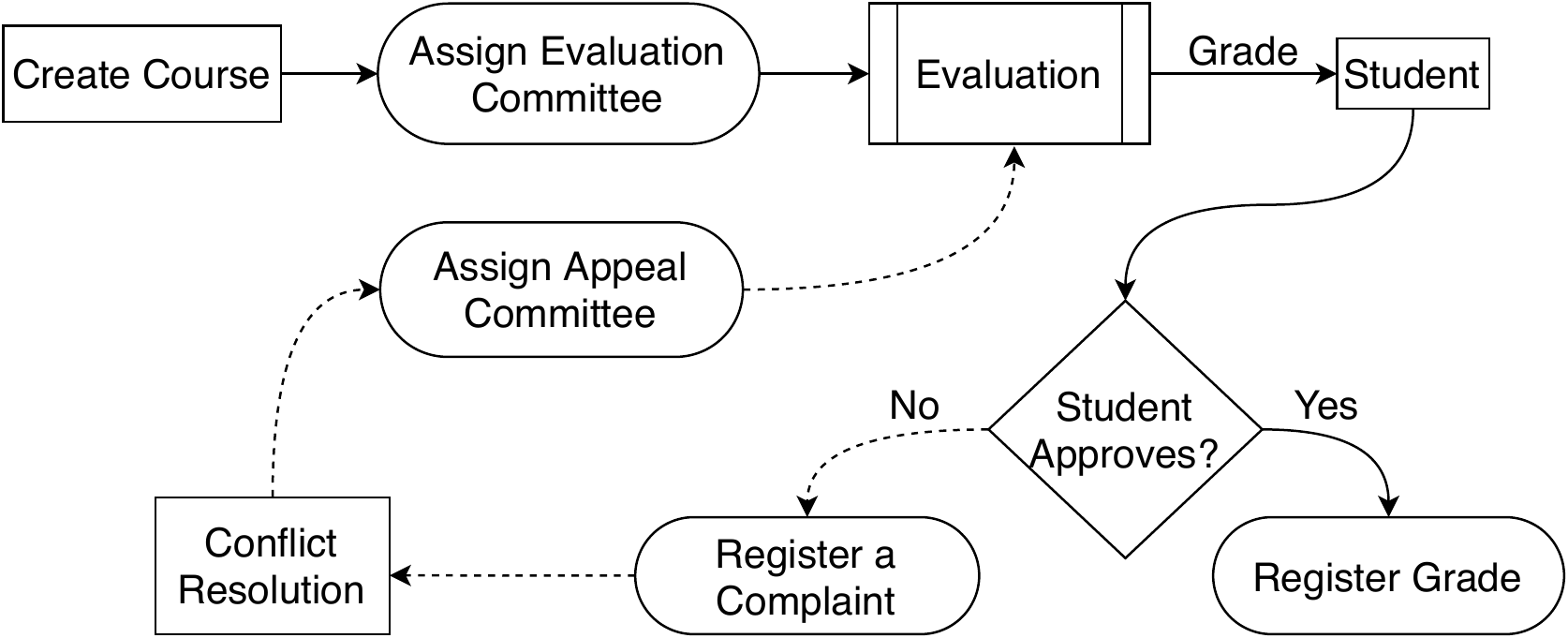}
      \caption{Example of an exam evaluation process which allows appeal.}
      \label{fig:evaluation-process-example}
      \vspace{-0.5cm}
\end{figure}

\noindent
Our protocol allows us to augment the certification processes with complex behaviors.
For instance, consider the evaluation process shown in \figurename~\ref{fig:evaluation-process-example}.
This example is representative of existing processes at many universities.

The process is modeled using our protocol by generating an exam \sdiploma, and its respective \emph{credential proofs}, in the course contract for each exam taken by each student.
Each \emph{exam proof} must be signed by all assigned evaluators and approved by the student to be considered valid.
Such authenticated operations are performed by calling the course contract.
The process repeats for the duration of a course.
At the end of the course, the evaluators aggregate the students' \sdiplomas and produce the credential root for each student, attesting to their final grades.

If a student does not agree with some grades, the process allows the student to request an appeal for a new evaluation with different evaluators.
Administrative staff manages this process by assigning an appeals committee to reassess a student's exam.
This appeals process can also be modeled using our protocol without adding extra functionality.
The administrative staff simply creates a new notary representing the new evaluation process under the department contract.

\subsection{Diploma Construction}

\noindent
A diploma is an \diploma whose credential tree is rooted in the university contract.
The diploma's tree comprises a set of \sdiploma subtrees successively aggregated during the certification period as described in Section~\ref{sec:credential-construction}.
The course contracts issue certificates representing individual exams, while department contracts issue transcripts of records, aggregating completed courses each semester.
Later, the university contract issues a single credential for a completed degree, aggregating records from the different department contracts.
Finally, the anchor contract issues public evidence of the student's diploma creation in the permissionless \blockchain.

\subsection{Extending a Diploma}
\noindent
Universities can extend credentials by expanding the credential tree of a student.
For example, suppose credential $P12$ in \figurename~\ref{fig:cert-tree} represents a bachelor's degree at university $F$.
This degree can be referenced in a credential tree for a master's degree, either at university $F$ or another university, by applying the composability mechanism.

\subsection{Verifying Credentials}
\label{sec:verifying-credentials}

\noindent
Verifying \diplomas with our protocol encompasses two main parts: verify the credentials' authenticity and verify that the respective credential tree was correctly constructed.

Verifying a credential's authenticity requires that the student shares a verifiable presentation with the verifier.
Given such credential data, a verifier first extracts the notary address from the credential and then \emph{ensures} that the:
\begin{enumerate}
      \item University's ENS entry matches the anchor's address.
      \item University's root contract address exists in the anchor.
      \item The public evidence in the anchor contract matches the metadata in the root notary of the university.
      \item Credential's hash matches the metadata in the notary of the university and is signed by the correct registrars.
      \item Student address in the credential matches the metadata in the tree and is indeed the correct holder.
      \item Credential has not expired or been revoked.
\end{enumerate}

\noindent
In addition, if the credential contains creation evidence, the verifier can perform a tree traversal over the credential tree of the student, repeating Steps~4-6 for all witnesses of the credential until there are no more witnesses to be checked or credentials to be verified.
This latter part ensures that the evidence matches the metadata in the witness notary and that this witness is part of the university's certification tree.

Besides the authenticity and integrity checks of credentials, our protocol also allows verifiers to check the whole certification process.
For instance, a verifier can check the authenticity and integrity of a student's transcript of records and only inspect the content if the student discloses it.

To further improve transparency in the processes, verifiers can perform various validation checks based on the on-chain metadata.
We highlight the following:

\begin{itemize}
	\item The issuing time of any credential.
      \item The duration of all certification processes.
      \item The dependency between credential subtrees and notaries.
	\item The code of notaries that encodes the certification.
      \item The list of revoked credentials.
      \item All notaries of a certification tree.
      \item All authorized registrars in specific periods.
\end{itemize}


\section{Evaluating Deployment Options}
\label{sec:eval}

\noindent
This section evaluates our approach described in Section~\ref{sec:impl}.
We compare our approach with BlockCerts and analyze the tradeoffs between alternative deployment options described in Section~\ref{sec:credential-construction}.

\subsection{Permissionless Deployment}

\noindent
In a permissionless deployment, an anchor contract is not necessary, and the ENS entry of the university can point to the root contract of their certification tree, representing the university identity.
Although this deployment has a non-negligible cost, such cost also presents limitations for malicious actors.

In Ethereum, the cost is based on the amount of data sent, the fees collected by the miners, and the cost of executing the transaction, which can trigger contract execution.
Hence, the cost of our certification system is the total cost of transactions and the execution cost of the notary code.
The cost also depends on the system demand and market variations.
For comparison, we assume a fixed Ether exchange rate.

Deploying our system on a permissionless \blockchain provides the best resistance against participant collusion and strong infrastructure reliability.
However, as our test scenario considers the current Ethereum implementation, which is based on proof-of-work, there is an inherent limitation on the latency and throughput of the system. For example, Ethereum is currently throughput-limited at approximately 15~tps.

While the academic credentials use case is not latency constrained, it would be problematic if an ever-increasing number of distributed applications run on the same platform, resulting in increasing latencies due to the throughput limitation.

BlockCerts is deployed in a permissionless \blockchain and currently supports Bitcoin and Ethereum, but does not use \contracts.
Their issuing process consists of sending a transaction with the credential's hash from the university's address to the subject's address.
Thus the cost is essentially the cost of a send transaction.
BlockCerts can also store the root of a Merkle tree of a batch of credentials.
However, the transaction cost remains the same, independent of the number of credentials in the batch since the root is a single hash.

Table~\ref{tab:costcomp} compares the cost of creating a credential for a student using our protocol and BlockCerts.
For the canonical BlockCerts approach, only the hash of the student's diploma is published.
Issuing such a credential is the cheapest transaction cost possible using Ethereum.
However, it assumes trust in the issuer and does not provide any publicly verifiable information on how the credential was derived.

It is also possible to use BlockCerts to maintain a progressive log of the student's achievements by issuing credentials more frequently, e.g., per exam completion.
This increases the cost linearly with the number of issued credentials.
Although this approach improves transparency at a reasonable cost, it does not prevent the issuance of fake credentials.
All credentials depend on the same signing key.

As described in Section~\ref{sec:impl}, issuing a credential in our protocol requires multiple interactions between users through contracts.
All actions are logged for accountability, resulting in a higher cost than issuing a single hash on the blockchain.
To create a valid diploma, university staff must collaborate in a progressive credential construction based on a set of authorization steps performed by the notary contracts.
Thus, university staff cannot abuse the certification system.

Table~\ref{tab:costcomp} shows the costs involved in creating credentials at different levels of a certification tree of a university deployed in a permissionless \blockchain.
For simplicity, we assume that all courses represent 10 ECTS and that a semester represents 30 ECTS.
Further, each course is assumed to have 5 exams, issuing a credential for each exam.
The cost of an exam in our permissionless approach is 11 times the cost of issuing a hash using BlockCerts.

\subsection{Hybrid Deployment}

\noindent
Our protocol can be deployed in a hybrid blockchain infrastructure to reduce operational costs and retain the transparency provided by the certification tree.
Nevertheless, such an approach assumes a certain level of trust in the institutions that maintain the system.

The certification tree is deployed in a permissioned \blockchain composed of a consortium of universities.
Each university is a validator of the \blockchain and can add new blocks.
Each university also deploys an anchor contract, described in Section~\ref{sec:anchor}, on a permissionless \blockchain.
The anchor is triggered periodically to publish the credential roots.

Using a permissioned \blockchain has better performance and cost benefits while preserving the immutability and time\-stamping properties of the \blockchain.
The anchors allow external entities to retrieve the credential root for a specific period of a university's certification tree that the university cannot manipulate after it has been published.
This root can be used to match the data in the permissioned \blockchain, ensuring the correctness of the certification process of each institution. The approach also allows on-chain revocation and transparency of the certification process.

Anchors on Ethereum are currently protected by proof-of-work. Thus, the integrity of credentials and the certification process can be validated even if the permissioned infrastructure is no longer deemed reliable or if cryptographic primitives are broken. We believe this to be a valuable feature for \diplomas.

The cost and storage requirement of this approach is shown as Hybrid in Table~\ref{tab:costcomp}.
The table shows the minimum amount of data stored on-chain in the anchor contract.
The storage requirement and cost of the permissioned \blockchain infra\-structure are not considered.

As Table~\ref{tab:costcomp} shows, the cost of a bachelor's diploma using our protocol in the hybrid deployment can be around 300 times cheaper than the same diploma in the permissionless deployment and 11 times more expensive than BlockCerts.
While this cost may seem high compared with the canonical BlockCerts approach, BlockCerts does not offer the same level of transparency, resilience, revocability of credentials, and flexibility that our protocol can offer.

Further, BlockCerts uses a minimum amount of on-chain storage since only a single hash is stored once a year or so.
On the other hand, deploying our protocol on a permissionless blockchain requires a non-negligible amount of on-chain data.
The data is mainly used to allow the \contracts to perform on-chain validation of the certification process, which cannot be done in the BlockCerts model.
In the hybrid deployment, such data can be kept in the permissioned \blockchain, allowing us to reduce the on-chain storage overhead of our protocol considerably, only publishing the credential roots in the anchor contract, similarly to what is done by BlockCerts.

\newcolumntype{C}{>{\centering\arraybackslash}m{1.2cm}}
\newcolumntype{L}{>{\raggedleft\arraybackslash}m{1.2cm}}

\begin{table}
\renewcommand{\arraystretch}{1.3}
\caption{Cost and On-chain Storage Comparison}
\label{tab:costcomp}
\begin{center}
\scalebox{0.8}{
\begin{tabular}{ l l | c | L | r | r }
\toprule
Protocol                &                    & ECTS & \multicolumn{1}{C|}{Credentials Issued} & Storage & Cost USD$^{*}$        \\ \midrule
\multicolumn{2}{l|}{BlockCerts Canonical}    &      & 1                                       &    \blockcertssize     &    \blockcertscost{1}            \\
\multicolumn{2}{l|}{BlockCerts Progressive}  & 180  & 90$^{\dagger}$                               &    \blockcertsprogsize{90}     &   \blockcertscost{90}            \\ \midrule
Permissionless          & Exam cost          &      & 1                                       &   \examcredsize     & \examcost{}           \\
                        & Course cost        & 10   & \issuedcred{5}                  &  \coursecredsize{5}    & \coursecost{5}        \\
                        & Semester cost      & 30   & \issuedcred{3*5+1}                  &  \semestercredsize{3}{5}    & \semestercost{3}{5}   \\
                        & Bachelor's diploma & 180  & \issuedcred{6*3*5+6+1}                  &  \diplomacredsize{6}{3}{5}    & \diplomacost{6}{3}{5} \\
                        & Master's diploma   & 120  & \issuedcred{4*3*5+4+1}                  & \diplomacredsize{4}{3}{5} & \diplomacost{4}{3}{5} \\ \midrule
Hybrid                  & Exam cost          &      & 1                                       &   -     & -           \\
                        & Course cost        & 10   & \issuedcred{5}                          &    -     & -        \\
                        & Semester cost      & 30   & \issuedcred{3*5+1}                      &    -     & -   \\
                        & Bachelor's diploma & 180  & \issuedcred{6*3*5+6+2}                  &  \examcredsize   & \examcost{}  \\
                        & Master's diploma   & 120  & \issuedcred{4*3*5+4+2}                  &  \examcredsize       & \examcost{} \\ \bottomrule
\multicolumn{6}{l}{$^{*}$ Gas price $= \gas$ Gwei and ETH $1 = \ethusd$ USD at \daterate.\quad $^{\dagger}$ One per exam.}                                      \\
\end{tabular}
}
\end{center}
\vspace{-0.8cm}
\end{table}

Our protocol could also be used to extend the BlockCerts threat model and use BlockCerts as an anchor. However, as BlockCerts does not use \contracts, the revocation of credentials is still dependent on an issuer-hosted revocation list, representing a single point of failure and trust.
Further, our protocol provides an automated and secure approach to manage the relationship between all credentials that compose such diplomas.
These are publicly verifiable and independent of the issuer's authorization, or even if the issuer institution no longer exists.


\section{Conclusion}
\label{sec:conclusion}

\noindent
In this work, we propose three mechanisms: credential composability, trusted timestamp, and distributed responsibility, that enhance the security of certification processes and digital credentials.
The mechanisms create a dependency graph of related credentials and enable accountability of the certification process.
We implemented these mechanisms to create credentials in a tree-like data structure with causal dependency properties.

Our construction can prevent several types of fraud by limiting the creation of invalid credentials, offering powerful tools for detecting bogus certification processes.
Further, we evaluate our protocol in two different deployment scenarios for the academic use case, illustrating complex certification procedures, such as issuing diplomas with strong transparency guarantees.

We believe that the transparency of the certification process will bring benefits for accountability, legitimacy, and trust for the institutions that use the system.
Our design will facilitate the creation of incentive mechanisms to inhibit fraud and help detect degree mills.

\section*{Acknowledgment}

\noindent
This work is partially funded by the BBChain and Credence projects under grants 274451 and 288126 from the Research Council of Norway.

\bibliographystyle{IEEEtran}
\bibliography{IEEEabrv,references}

\end{document}